\documentclass[10pt]{article}
\usepackage{amssymb,amsfonts,amsmath}
\usepackage{graphicx}
\usepackage{cite}
\usepackage{color}
\usepackage{array}

\usepackage[labelfont=bf,labelsep=period,justification=raggedright]{caption}

\makeatletter
\renewcommand{\@biblabel}[1]{\quad#1.}
\makeatother

\date{}

\begin{document}

\begin{flushleft}
{\Large
\textbf{Neutral selection}
}
\\
David A. Kessler$^{1,\ast}$.
Nadav M. Shnerb$^{1}$\\
\bf{1} Department of Physics, Bar-Ilan University, Ramat-Gan IL52900, Israel
\\
$\ast$ E-mail: kessler@dave.ph.biu.ac.il
\end{flushleft}

\section*{Abstract}

Hubbell's neutral theory of biodiversity has successfully explained
the observed composition of many ecological communities but it
relies on strict demographic equivalence among species and provides
no room for evolutionary processes like selection, adaptation and
speciation. Here we show how to embed  the neutral theory within the
Darwinian framework. In a fitness landscape with a quadratic maximum,
typical of quantitative traits, selection restricts the extant
species to have traits close to optimal, so that the fitness
differences between surviving species are small.  For sufficiently
small mutation steps, the community structure fits perfectly to the
Fisher log-series species abundance distribution. The theory is
relatively insensitive to moderate amounts of environmental noise,
wherein the location of the fitness maximum changes by amounts of
order the width of the noise-free distribution.   Adding very large
environmental noise to the model qualitatively changes the abundance
distributions, converting the exponential fall-off of large species
to a power-law decay, typical of a neutral model with environmental
noise.

\section*{Introduction}

One of the major challenges of modern ecology is to understand the
mechanisms that govern empirically observed patterns of species
richness and species abundance. Of particular importance is the
species abundance distribution (SAD), the relative proportions of
frequent versus rare species in an ecocommunity. In many cases this
curve  exhibits the coexistence of many species, all belong to the
same trophic level, within a relatively small area. This poses two
long standing puzzles. First, one has to explain the apparent
violation of the competitive exclusion
principle~\cite{hardin,tilman_book} that sets the scene for
evolution by means of natural selection~\cite{darwin}. Second, we
would like to understand the origin of the specific characteristic  shape of the SAD
curve measured for many different communities.

The solutions suggested to the biodiversity puzzle  range between
two extremes: the niche and the neutral.  Niche models
\cite{mc,may,clark2010} propose that each species has its own niche
with regard to the ensemble of resources used by the community, or
alternatively that a given species cannot fully exploit its niche
since its colonization ability is limited~\cite{tilman}. The
competitive exclusion principle then only applies within a given
niche, but competition between species in different niches is
limited. Mathematically speaking, it is assumed that the
deterministic dynamics of the system supports an attractive manifold
(equilibrium point or limit cycle) with many coexisting
species~\cite{Gross}. Neutral theories~\cite{book,maritan1,leigh,TREE2011}, on the other hand, suggest that all species are
demographically equal, and that the relative abundance of a species
is governed by a stochastic birth-death process with slight effects
of migration (in the case of local communities) or mutations (for
metacommunities).

Neutral theories were first proposed in the context of molecular
biology, where ``species" are different DNA sequences that code for
different proteins, and, following the works of Hubbell, became
extremely popular (although hotly debated~\cite{mcgill}) in the
contemporary ecological literature~\cite{TREE2011}. The assumption
of neutrality solves the coexistence problem, and the simple
community drift process indeed explains the empirically observed
species abundance distributions (SADs) in many ecocommunities, from
the tropical forest to  coral reefs~\cite{reef,he}.

Conceptually, however, the  neutral approach has raised a severe
problems. Community ecology is supposed to be the stage on which the
evolutionary play unfolds. If the community dynamics is governed
solely by chance, there is no room for natural selection. Even
worse, it seems that there is no way to include weak selection
within the neutral framework. Indeed, it has been shown that any
deviation from the assumption of perfect demographic equivalence
leads to fixation in the community of the fittest
species~\cite{weak2,weak1}. Recent "continuum" models that suggest a
single framework for both  niche and neutral dynamics are focused on
the case of a neutral dynamics of many species within different
niches, with or without an overlap~\cite{kadmon,gravel,pacala,
zillio}. These models do not allow for evolutionary dynamics and in
fact species that belong to non-overlapping niches do not interact.

A framework that incorporates evolution and some features of neutral
dynamics was suggested in~\cite{Scheffer1,Scheffer2}. These authors,
however, implement a non-blind version of "evolution": only
mutations that changes the relevant trait in the direction that
leads to less competition are allowed. We will return to this subject in the discussion.


\begin{table}[!h]
\centering
\begin{tabular}{|c| m{3.2cm} | m{3.2cm} | m{4cm} | m{3.5cm}|} \hline
1 & \parbox[c]{1em}{\includegraphics[width=3cm]{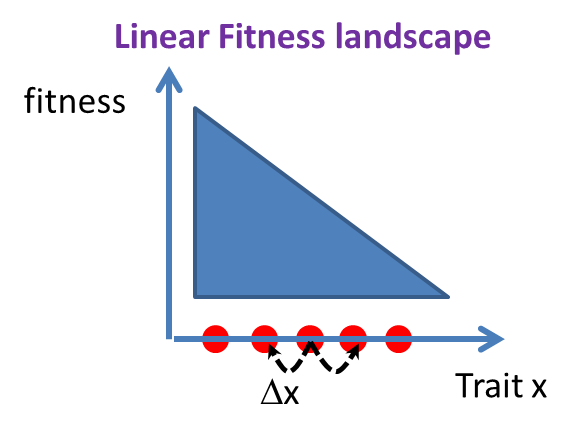}}
\label{lin}&
\parbox[c]{1em}{\includegraphics[width=3cm]{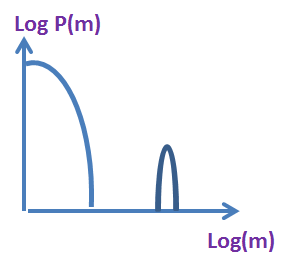}}
\label{Dardom}&  \parbox[c]{1em}{\includegraphics[width=3.8cm]{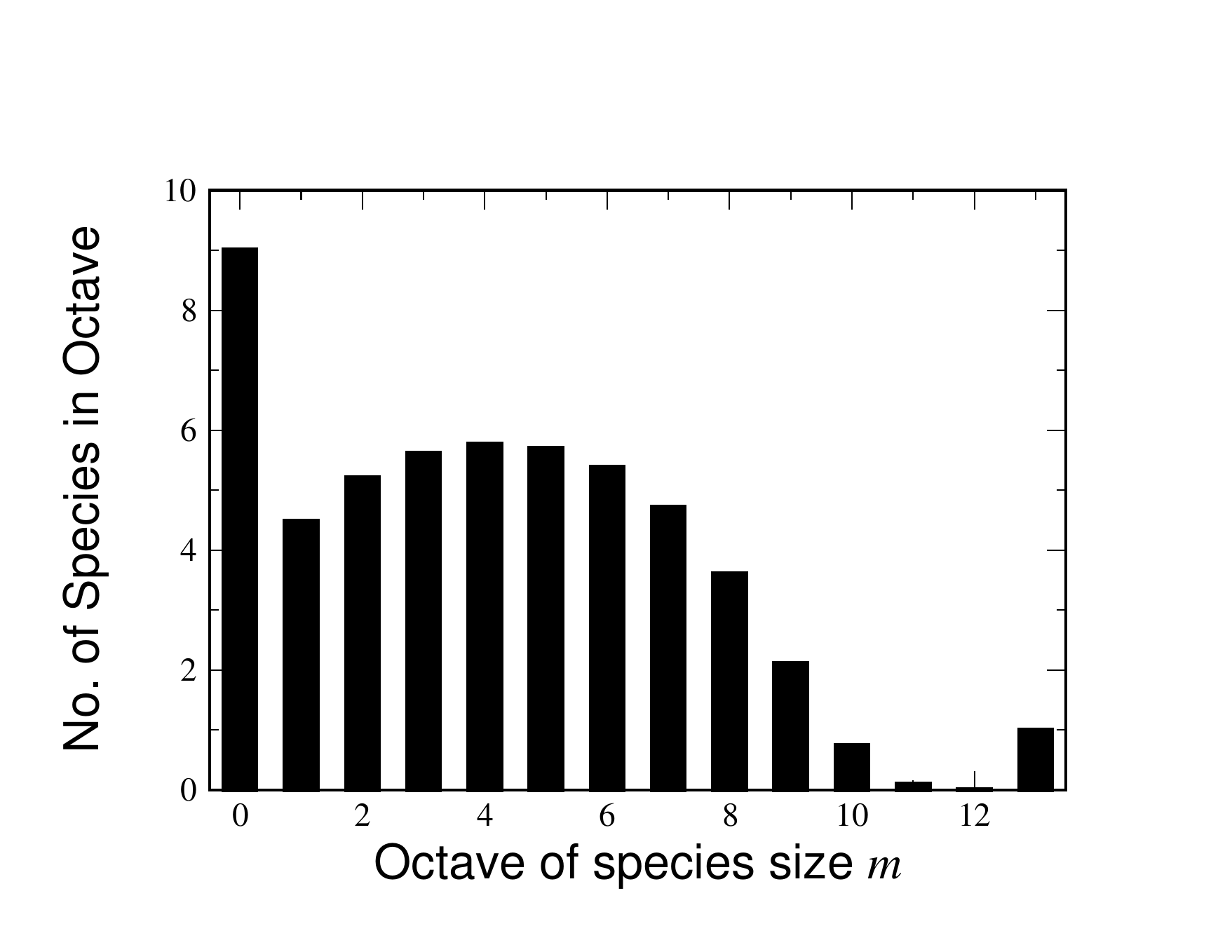}}
\label{prestln}&  \emph{Darwinian dominance}: the strongest species dominates.
\\ \hline
 2 &   \parbox[c]{1em}{\includegraphics[width=3cm]{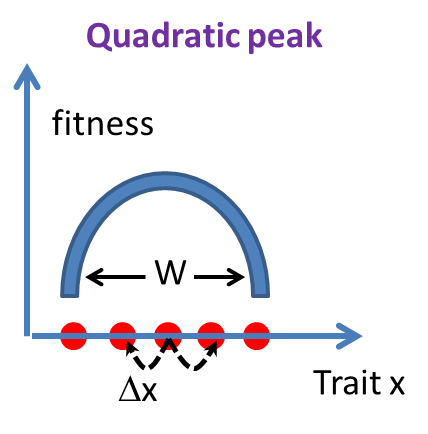}}
\label{quad}& \parbox[c]{1em}{\includegraphics[width=3.cm]{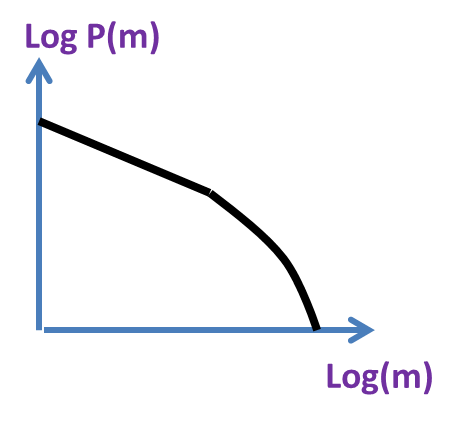}}
\label{Fisher} & \parbox[c]{1em}{\includegraphics[width=3.8cm]{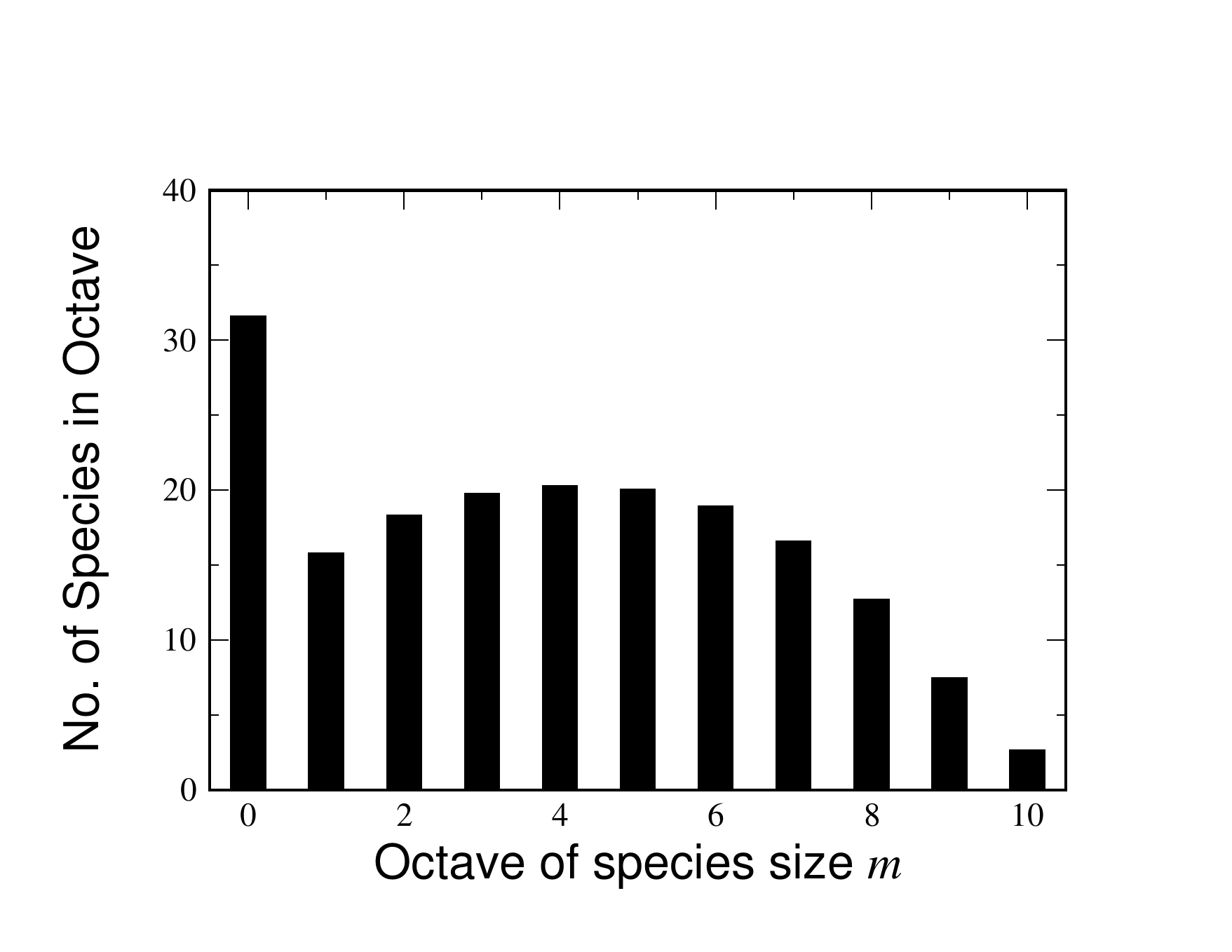}}
\label{prestfls}&  Generalized Fisher log-series Eq. (\ref{gfls}). $\bullet$ Strictly neutral (Hubbell-Kimura, $\beta = 1$). $\bullet$ Quadratic fitness peak without (or with weak) environmental stochasticity. $\bullet$ Hamilton-May \emph{with} local catastrophes.
  \\ \hline

3 & \parbox[c]{1em}{\includegraphics[width=3cm]{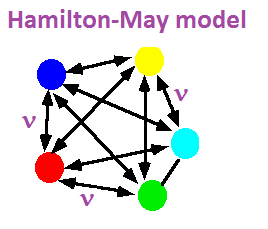}}
\label{HM} &
\parbox[c]{1em}{\includegraphics[width=3cm]{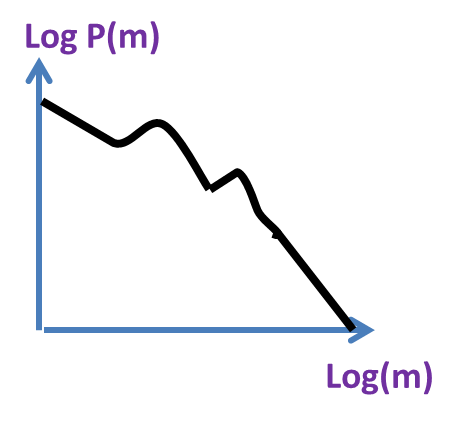}}
\label{HMniche}& \parbox[c]{1em}{\includegraphics[width=3.8cm]{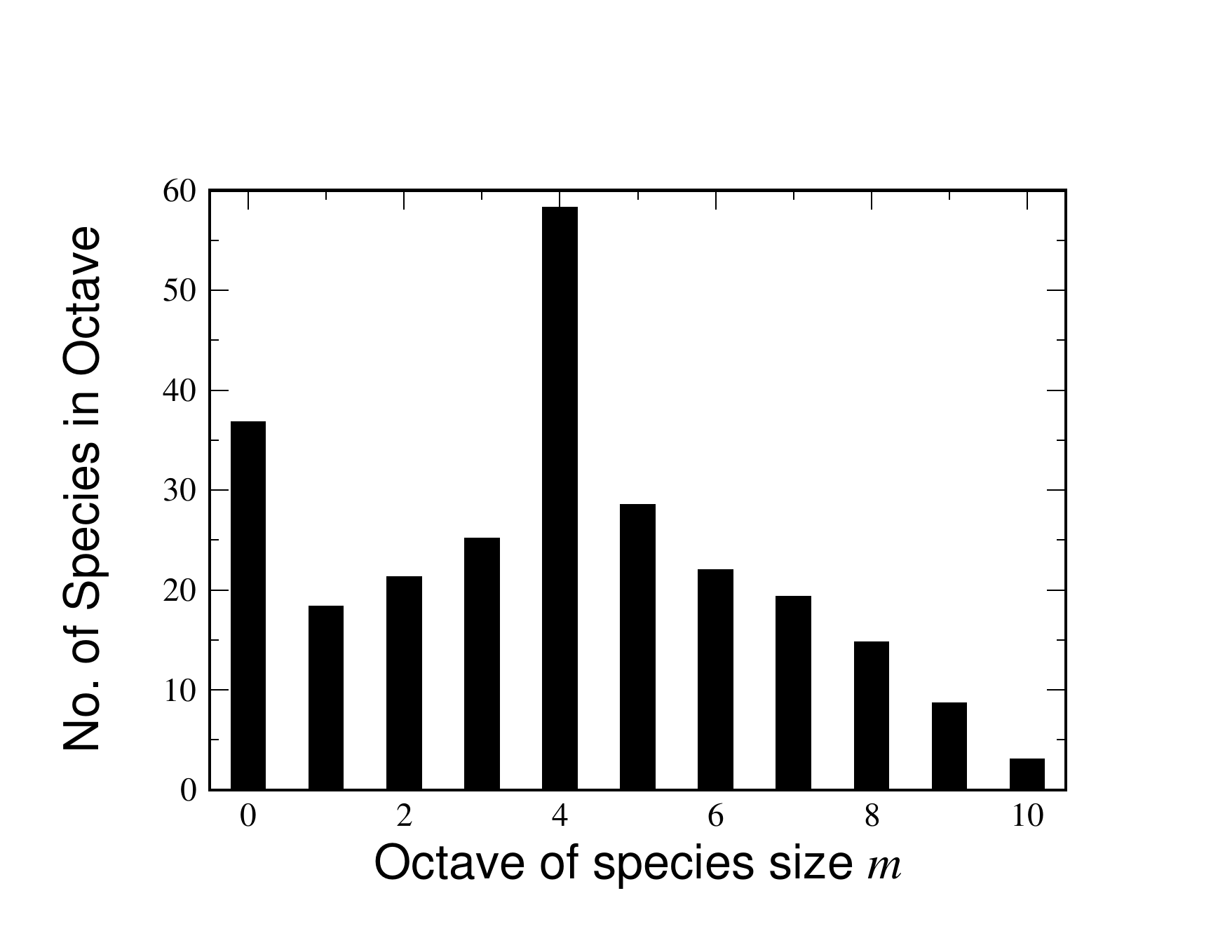}}
\label{prestniche}&  Hamilton-May model \emph{without} catastrophes \\ \hline

 4 &  \parbox[c]{1em}{\includegraphics[width=3cm]{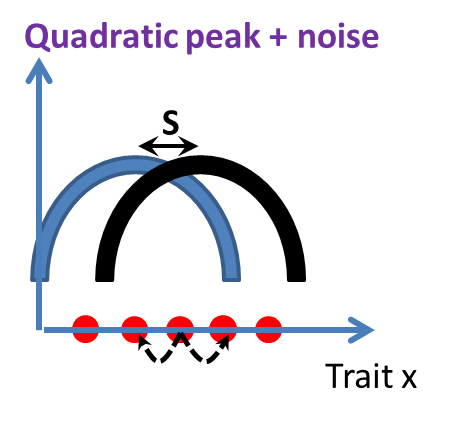}}
\label{quad+noise} & \parbox[c]{1em}{\includegraphics[width=3cm]{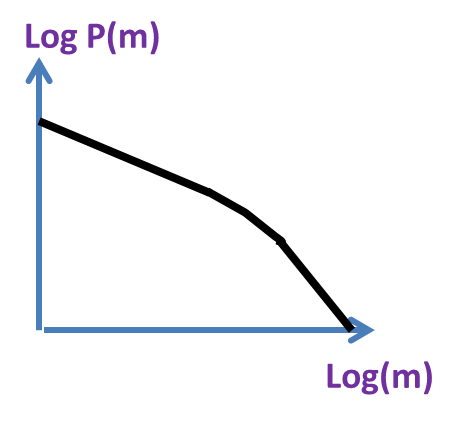}}
\label{envnoise}& \parbox[c]{1em}{\includegraphics[width=3.8cm]{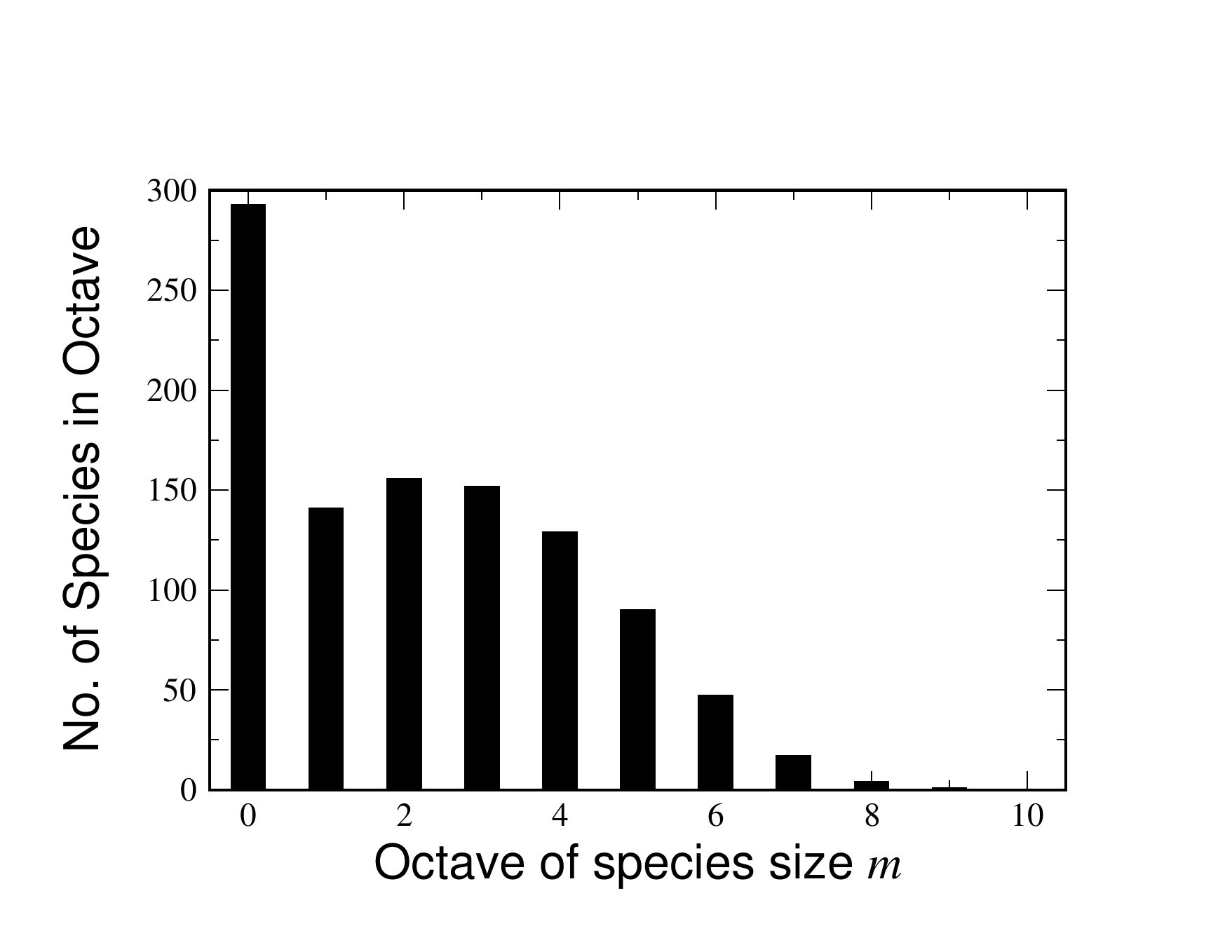}}
\label{prestenv}& $\bullet$ If $S/W  \ll 1$ Generalized Fisher log-series.
$S/W \gg 1$ leads to two power laws. $\bullet$  Neutral model with environmental noise, Eq. (\ref{ssnoise}). \\ \hline
 \end{tabular}
\caption{A qualitative summary of the SADs considered along this paper. Every line presents cartoons of one SAD genre, showing the normalized pdf on a double logarithmic scale (third column) and the corresponding Preston plot (fourth column). In the last column we supply a list of the models/cases  that yield this statistics, one of them is illustrated in the second column.    } \label{table1}
\end{table}



Here we will show  that the neutral theory of biodiversity may live
happily within  the Darwinian framework of species evolution via
natural selection, including blind mutations. The basic idea is
simple: when the fitness landscape is quadratic, traits that are
close enough to the peak have very similar fitness. Species with
really harmful traits will disappear, and those who survive will
play an almost neutral game.

The variety of SADs considered in this paper are classified, and illustrated by an appropriate cartoons, in Table 1. The first line corresponds to the "classical" example of Darwinian dominance, when the fitness landscape is linear. This case is considered in the first subsection of the results below. The second line illustrates the (Generalized) Fisher log-series, as explained in the second subsection of the results. In the third line and the corresponding
subsection we consider the classical  Hamilton-May model for evolution of dispersal rate and show that indeed this model may support a Fisher-like SAD in the presence of local  catastrophes, while in the absence of catastrophes it shows strong effect of niche, superimposed on the continuum SAD. Finally,  we discuss the case where SAD is a composition of two power laws.

The graphical presentation of SADs is subject to two common customs. Ecologists tend to implement Preston plots, showing the number of species for every octave of species size. The columns in such a histogram are not "normalized", since the actual probability that the abundance of a  randomly chosen species is within this octave, $P(m)$, is not the height of the column, but its height divided by the length of the relevant octave. In general, however, the standard way to present  power-law histograms is, of course, to plot $P(m)$ vs $m$ using (since the distribution is wide) logarithmic binning and a double logarithmic plot. We have found this second presentation more informative and applied it along this paper; however, to allow for a qualitative comparison with Preston plots, we present them in Table 1 for the SADs considered here.

\section*{Results}

In this section we present the results of numerical simulations, and some analytic calculations, regarding the models considered above. The details of the numerical procedure are given in the methods section below.

\begin{center}
\bf{Neutral SADs}
\end{center}

\begin{figure}
\begin{center}
\includegraphics[width=0.55\textwidth]{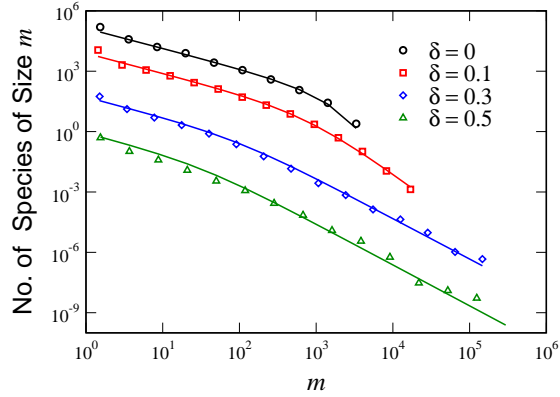}
\end{center}
\caption{Time-averaged SAD for the ``neutral environmental noise" model, for different values of the noise parameter $\delta=0$, $0.1$, $0.3$ and $0.5$.  The population size was $N=10^6$, and the mutation rate was $\mu=0.001$.  The fitness of each species was updated after $N$ pairwise competitions. The data for the case $\delta=0.1$ was fit to the formula Eq. (\ref{ssnoise}), with the line indicating the fit, given by $\sigma^2/(2S)=850$,
$\mu/S=.850$, consistent with $\mu=0.001$, $\sigma^2=2$.  The other lines were determined using the same fit values of $\mu$ and $\sigma^2$, and $S$ given by 0, $9S_{0.1}$ and $25S_{0.1}$ respectively, since the noise amplitude should scale as $\delta^2$. The graphs for the cases $\delta=0.1$ and $0.5$ were shifted up and down for clarity.}
\label{figenvneutSAD}
\end{figure}

To set the general framework for the analysis, let us present three relevant SADs:

\begin{itemize}
  \item The most famous neutral SAD emerges from the  Hubbell-Kimura community-drift model. If every individual has, on average, one offspring in the next generation, the noise that drives the system is pure demographic stochasticity and its strength is determined  by $\sigma^2$, the variance of the number of offspring per individual~\cite{maruvka3}. An  offspring belongs to the same species with probability $1-\mu$ and originates a new species with probability $\mu$. the SAD (the chance, $P(m)$ that a species  picked at random from the list of all existing species has abundance $m$) depends on these two parameters and  is given by the Fisher log-series (FLS)
      \begin{equation} \label{fls}
      P(m) = \frac{e^{-\frac{2 \mu m}{\sigma^2}}}{m}
      \end{equation}
  \item A generalized form of FLS is
  \begin{equation} \label{gfls}
      P(m) = \frac{e^{-\alpha m }}{m^\beta}.
      \end{equation}
      While this formula is not an analytic result of any neutral model we are familiar with, it clearly converges to (\ref{fls}) in the limit $\beta \to 1$. If $\beta$ is not too far from unity it is quite hard to tell apart (\ref{gfls}) from (\ref{fls}) given the limitations of currently available empirical datasets.
      
  \item The neutral model may be generalized to include not only demographic noise but also environmental stochasticity. This is a generalization of the concept of neutrality~\cite{alonso2007}: species do respond differently to environmental fluctuations, and their fitness varies correspondingly, but their fitness is equal to that of other species after averaging over a sufficiently long period of time. This introduces a new parameter to the theory: $S$, the strength of environmental noise. In the methods section we will show that, in this case,
      \begin{equation}
P(m) = \frac{A}{m} \left(m + \frac{\sigma^2}{2S}\right)^{-(1+\mu/S)}.
\label{ssnoise}
\end{equation}
      This result is a superposition of two power-low decays, the one characterizing the small $m$ behavior with exponent 1 as in Fisher and a second for $m \gg \sigma^2/(2S)$, with the larger exponent $1+\nu/S$.  For weak environmental noise, the second exponent diverges and the result is the standard Fisher log-series, with its exponential decay at large $m$. This analysis is confirmed by a simulation of the above environmental noisy neutral model, as shown in Fig.   \ref{figenvneutSAD} in the methods section below.
\end{itemize}

\subsection*{Darwinian dominance}

\begin{figure}
\begin{center}
\includegraphics[width=0.55\textwidth]{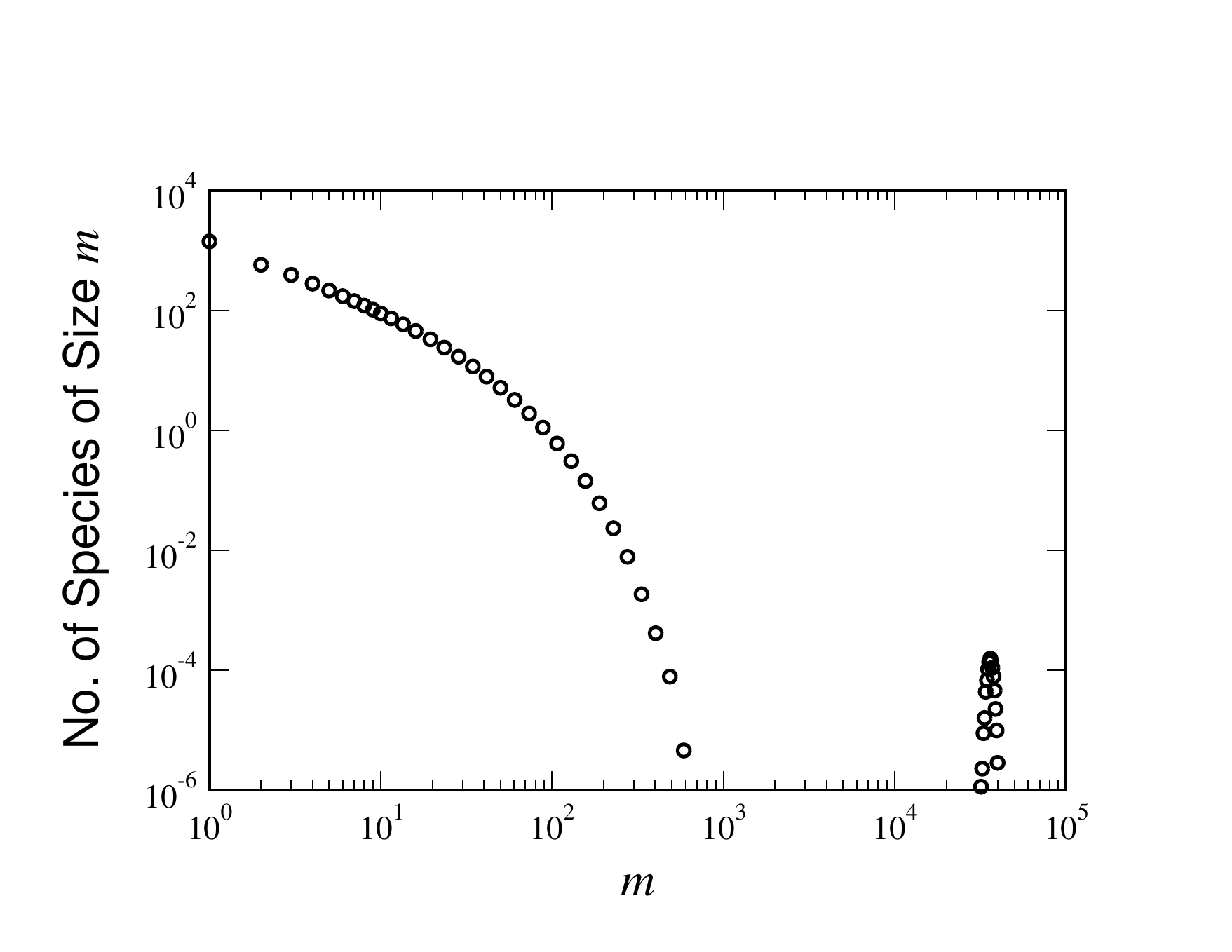}
\end{center}
\caption{Time-averaged SAD for the linear landscape with $s=0.01$, $\mu=0.01$, $N=10^5$. Notice the presence of the isolated peak at high $m$, representing the maximal fitness species, with approximately $35\%$ of the entire population, almost a factor of 10 larger than the next largest species. Here the fitness of the 2nd most fit state corresponds exactly to the fitness of the 2nd most fit state in the quadratic landscape with the strongest selection shown in Fig. \ref{figmanysSAD}.  Ignoring the isolated peak, the rest of the SAD curve may be fit to the generalized Fisher log-series, Eq. (\ref{gfls}), with $\alpha=0.019$, $\beta=1.25$. The normalization of the generalized Fisher log series is off however, reflecting the fact that a third of the population, contained in the most fit species, is unaccounted far.}
\label{linSAD}
\end{figure}

  Line 1 of Table 1 illustrates the traditional Darwinian scenario of fitness landscape, where fitness is considered as a linear function of some trait $x$. Every species (or haplotype, or phenotype) admits a single value of $x$ (red circles), and at any moment the species with the lowest $x$ selects out all other species. If individuals may undergo mutation and change their trait by $\Delta x$, the overall fitness of the population will improve over time until it reaches the maximum fitness at the endpoint. Although the competitive exclusion principle suggests that the only type that will survive is the fittest, this is strictly true only when the rate of mutations approaches zero. For finite mutation rate one expects a "cloud" of low-abundance species together with   high-abundance  one - the fittest.  Figure \ref{linSAD}, which corresponds to the cartoon in the first line of Table 1,  shows the SAD in that case as obtained from a standard Wright-Fisher simulation (see methods).  There is a large ratio between the population of this optimal species and that of the next most populous one.  This is reflected clearly in the figure, where there is an isolated peak in the SAD lying far to the right of the rest of the curve,  corresponding to the population distribution of the unique fittest species.  While the rest of the curve  does fit a generalized Fisher log-series, Eq.  (\ref{gfls}), the isolated peak (which in a single time snapshot would be an isolated outlier point with very high population) shows that this case is easily distinguished from the neutral case. 
  
Furthermore, there is an essential difference between the behavior of the single peak associated with the dominant species and the rest of the SAD under change of population size.  If the population is doubled, the form of the SAD (except for the isolated peak) remains the same, with the number of species of each size doubling. The isolated peak, however,  remains a single species, of {\em twice the size}. Looking at a subsample of half the size, then, is a easy way to test for a linear-type landscape.

\subsection*{Quadratic fitness landscape}

\begin{figure}
\begin{center}
\includegraphics[width=0.55\textwidth]{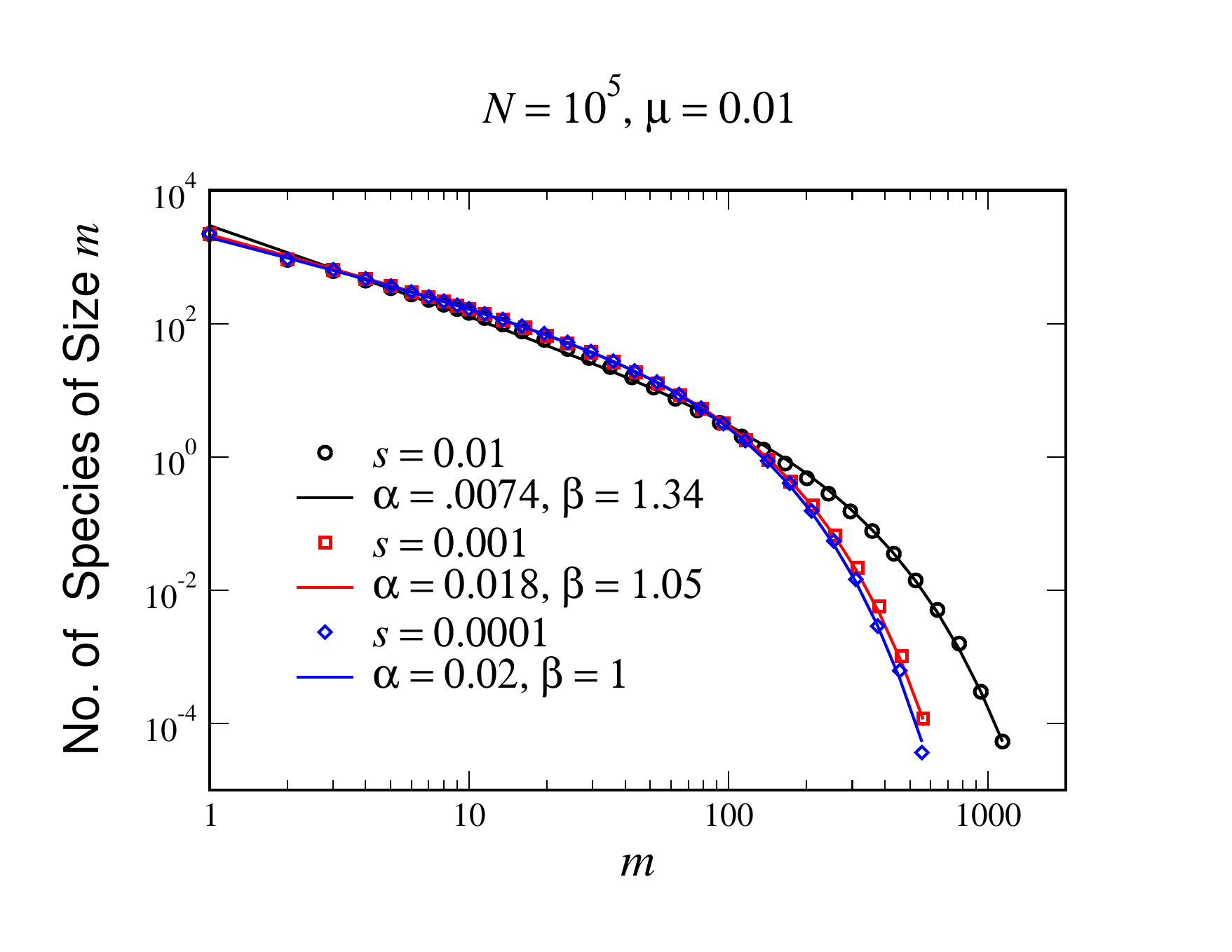}
\end{center}
\caption{Time-averaged species abundance curve for the quadratic landscape showing the mean number of species of size $m$ vs. $m$, for different values of the selection strength $W$, with fixed mutation step $\Delta x=\pm 1$. Here the population size $N=10^5$ and the mutation rate is $\mu=0.01$.  Also shown are fits to the generalized Fisher log-series, Eq. (\ref{gfls}) for each of the data sets.  As $W$ decreases, the exponent $\beta$ approaches unity.}
\label{figmanysSAD}
\end{figure}

\begin{figure}
\begin{center}
\includegraphics[width=0.55\textwidth]{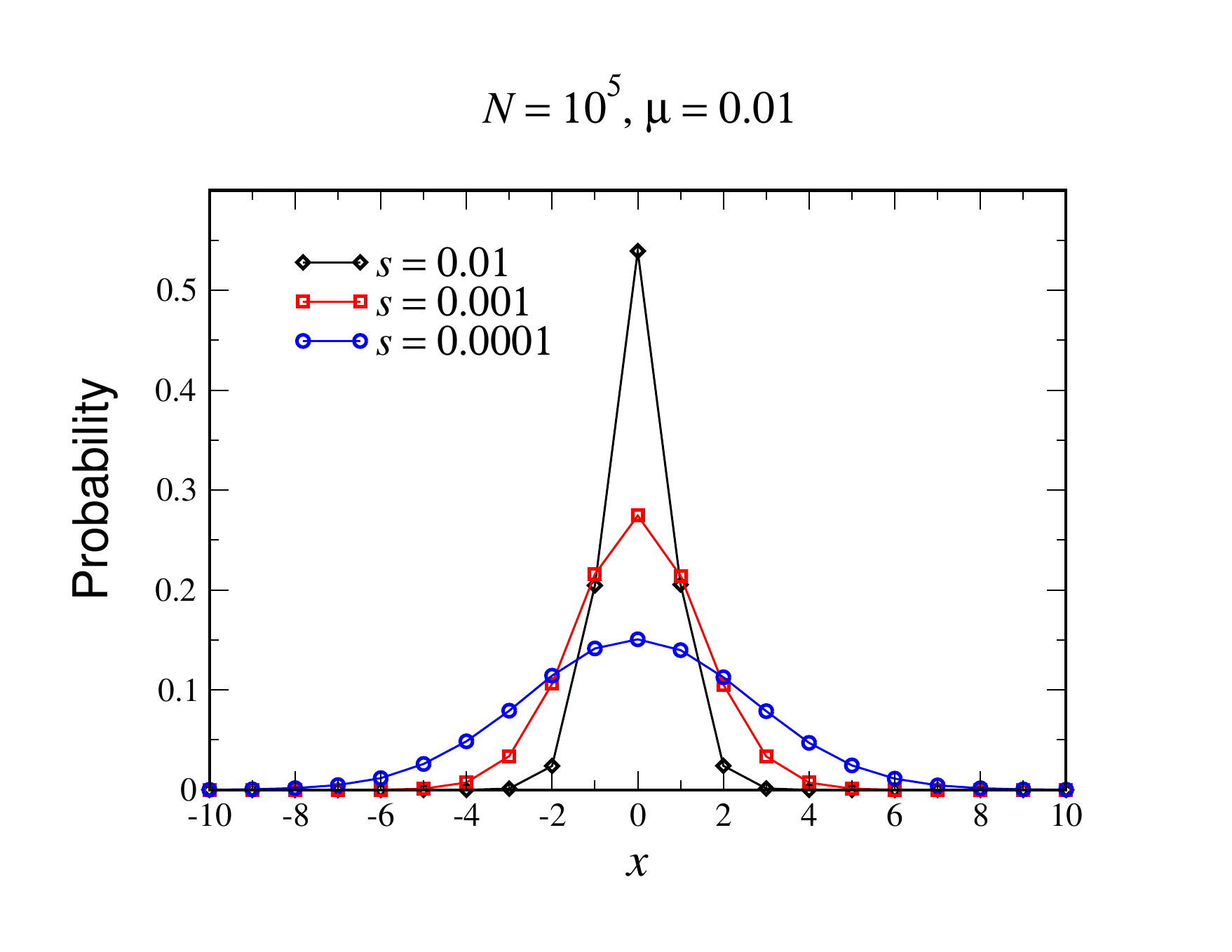}
\end{center}
\caption{Time-averaged probability distribution of $x$ for the quadratic landscape, with the same parameters as in Fig. \ref{figmanysSAD}, showing that selection ensures that the population is strongly peaked around the fitness maximum, $x=0$.  Decreasing $W$ widens the distribution, but in all cases the range of $x$ is quite limited. The lines are drawn simply to help visualize the data.}
\label{figmanysPx}
\end{figure}

The second line of Table 1 describes a fixed quadratic landscape, with a
maximum at $x=0$ and width $W$.   Identifying
each mutated individual as the founder of  a new ``species", we calculated the
species abundance distribution (SAD) presented in Fig.
\ref{figmanysSAD}.  We also measured the distribution of $x$ in the
metapopulation, as shown in Fig. \ref{figmanysPx}. As expected, the
$x$ distribution was strongly peaked about the maximum, whereas the
SAD  looked quite similar to the Fisher log-series distribution of
the neutral Hubbell model. Indeed, the SAD fits extremely well to a
generalized Fisher log-series given by Eq. (\ref{gfls}). Moreover, the exponent $\beta$ was found to decrease with the size of the mutation step (relative to the selection strength), such that in the limit of small mutation step, $\beta$ approaches one and the original Fisher log-series is recovered. The mutation \emph{rate}, on the other hand, mainly affects the parameter $\alpha$, with only a weak impact on $\beta$ (Fig. \ref{figmanymu}).

A small $m$ power-law exponent different from unity was also seen in the model of Zillio and Condit~\cite{zillio}, and in fact the above model is a simplified limit of the model considered therein (see Discussion for more detail).

We note that it is essential that the optimal state be degenerate or very nearly so.  This is the case for our model with a quadratic maximum, since many different genes contribute to the trait $x$ and so there are many different ways of achieving optimality.

\subsection*{Hamilton-May model}

Our model of a quadratic landscape can be criticized as ad-hoc, imposed as it is by fiat.  To study a richer model where the fitness landscape arises out of the intrinsic dynamics, we have therefore investigated a second model, based on the Hamilton-May~\cite{hm} dispersal dynamics.  Here the hereditary trait under evolutionary control is the probability of dispersal, $\nu$, of an individual to a distant island. The dispersal  is assumed to carry with it a fixed cost in the form of a given probability to not survive the dispersal. This model possesses, in the limit of an infinite number of islands, an evolutionary stable strategy, $\nu^*$~\cite{hm, hm1}, which is guaranteed to outcompete any other value of $\nu$ and take over the entire metapopulation, in accord with the competitive exclusion principle.  This evolutionary stable strategy represents the optimal tradeoff between the costs of dispersal on the one hand and the benefits it entails due to the reduction of fluctuations, on the other. Thus, qualitatively the model behaves as if there is an effective (density-dependent) fitness with a quadratic maximum at $\nu^*$, even though quantitatively the effective fitness depends on the exact composition of the community, (i.e., the entire range of strategies deployed by the members of the community at any given time) and is not a property just of the individual's $\nu$ alone.  In our model, we allow infrequent mutations to change the $\nu$ of the offspring. In such a case, the community will at long times exhibit a narrow range of $\nu$'s centered around $\nu^*$ (Fig. \ref{fighmnuhist}).  Nevertheless, as with our observations on the fixed quadratic landscape, the species abundance distribution (identifying different $\nu$'s as belonging to different species) takes the form of the generalized Fisher log-series, Eq. (\ref{gfls}), with $\beta$ approaching 1 as the size of the mutational jumps in $\nu$ is reduced.

\begin{figure}
\begin{center}
 \includegraphics[width=.55\textwidth]{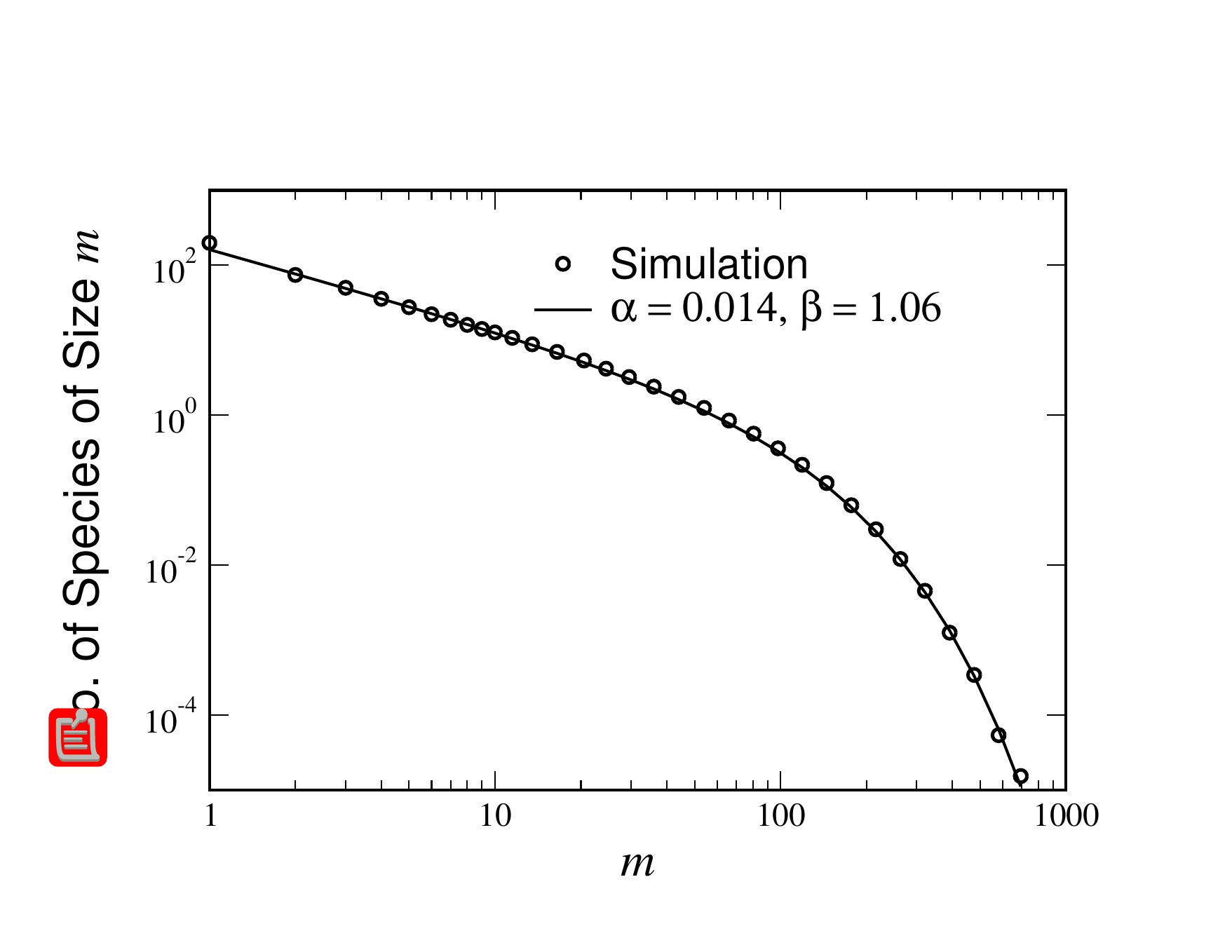}
\end{center}
\caption{Time-averages SAD for the Hamilton-May model, with 500 islands of capacity 20.  The dispersal survival probability, $p=0.3$, and the probability per generation for any given island to suffer  a catastrophe that wipes out the local population is $X=0.1$.  The mutation probability is $\mu=0.01$, and the parameter governing the mutation distribution is $\alpha=64$. Also show is the best fit to the generalized Fisher log-series, Eq. (\ref{gfls}).}
\label{HMSAD}
\end{figure}

It is interesting to note that the Hamilton-May model \textbf{without local catastrophes} exhibited a different behavior due to its spatial structure.
Here, many of the local communities were occupied by a
single species (see supplemental Fig. \ref{fighm0x}, Third line of Table 1). Essentially, each island constitutes in some measure a single niche which is shielded from the effects of competition from outside. The local catastrophes ensure that such niches do not last very long, and open up the island to competition from outside.

The Fisher log-series behavior of the Hamilton-May dynamics with local catastrophes is somewhat surprising given the large amount
of environmental noise in the model.  This noise manifests itself in two ways.  Firstly, the random local catastrophes on the islands is clearly a source of
environmental noise. For example, if the simulation utilizes L=500 islands, with 10\% chance of a local catastrophe, the number of local extinctions will vary around 50 with typical fluctuations of $\sqrt{50}$, and the optimal $\nu$ fluctuates as  well.   Secondly, the fact that fitness in the Hamilton-May model is dynamic, depending on the makeup of the entire metacommunity (the range of $\nu$'s in the population and their distribution among the various islands) means that each species sees a time-varying landscape.  This noisiness of the system is apparent when comparing the time-averaged probability density for $\nu$ with a single snapshot (Fig. \ref{fighmnuhist}). Nevertheless, the resulting SAD is apparently insensitive to this noise.

\subsection*{Quadratic fitness with environmental stochasticity}

\begin{figure}
\begin{center}
\includegraphics[width=0.55\textwidth]{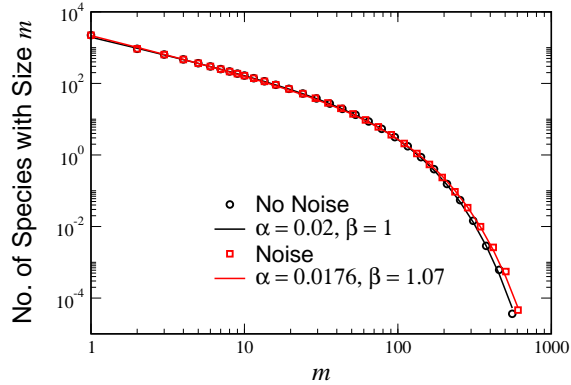}
\end{center}
\caption{Time-averaged SAD for the quadratic landscape comparing the case of environmental noise to the noise-free result. The parameters are as in Rig. \ref{figmanysSAD} with $s=0.0001$.  The parameters of the noisy case are $\tau=200$ and $\delta=0.5$.}
\label{figquadnoiseSAD}
\end{figure}

To put this finding in perspective, we explicitly added environmental noise to our quadratic landscape model, letting the $x$ of the fitness maximum fluctuate randomly in time around zero. This introduces another parameter, which is the amplitude $S$ of the optimal-fitness fluctuations (see the fourth line of Table 1), to be compared with the curvature parameter $W$. In the limit $S/W \to 0$ the environmental noise is negligible and the SAD is close to the Generalized FLS (\ref{gfls}).    Even when the noise was substantial  ($S$ of order of $W$) (see Fig. \ref{figxvst}), the SAD was hardly affected, as seen in Fig. \ref{figquadnoiseSAD}.

This behavior is in contrast to that expected from a ``neutral" theory with environmental noise. Here the neutrality is in a time-averaged sense, so that while there are fitness differences between species, they are random and fleeting. Indeed if we increase the noise further, $S/W \gg 1$, we do begin to see the impact of the noise on the SAD and the functional form becomes that of Eq. (\ref{ssnoise}).  This is seen in the supplemental figure, \ref{figquadnoises}.

\section*{Summary and Discussion}

We have presented an evolutionary model that reconciles Hubbell's neutral
model with selection, exhibiting the species abundance statistics of
the neutral model, even though selection acts on the quantitative
trait that characterizes different species in our model.  We have demonstrated this in two models, one with an imposed quadratic fitness landscape
and the other with a density-dependent fitness arising from the intrinsic dynamics of the Hamilton-May dispersal model.

The reason the model is able to circumvent the competitive exclusion
principle is that there are many species with essentially the same
fitness, plus a drift from the stronger competitor to other species due to mutation.  This is not a special feature of our model.  Any time
selection acts on a quantitative trait, we should expect that the
optimal value of the trait is somewhere in the middle of the
possible range, rather than being at an endpoint, as the standard
linear landscape picture~\cite{herbie1,herbie2} would suggest.  Indeed, a model with a single fitness peak at the endpoint
of a linear landscape shows this single species dominates the community.
In a quadratic landscape, however,  there is effectively a great deal of neutrality of species
in the vicinity of the  fitness maximum.

It should be noted that there is a nontrivial consistency condition
at work here. The evolutionary dynamics itself sets the width of the
fitness distribution of the community, and thus the degree of
(non)neutrality.  The weaker the selection, the wider the
distribution of traits in the community will be, and those species
at the fringe of survival will still be significantly less fit than
their compatriots with near optimal traits. The models presented above support coexistence without invoking too
strong  environmental noise to wipe out the details of the fitness
landscape.

In agreement with this rationale, it turns out that the size of the mutational step also plays a role.  When the mutational step is large, a generalized Fisher log-series emerges,
where the small species size power-law is changed from unity to a larger value. This was true both in the quadratic landscape model and in the Hamilton-May model.

As mentioned in the body, our quadratic landscape model is a limiting case of the model considered by Zillio and Condit~\cite{zillio}. They considered a spatially explicit model where the fitness maximum varied with location.  Our model is essentially equivalent to theirs in the case of a single ``niche". In some sense, their spatially dependent fitness is a kind of environmental noise, and indeed it appears that the large species behavior is power-law rather than exponential, consistent with our findings.  Since their data was not binned, however, it is hard to be more definitive at this stage.  It would be interesting to pursue this line of inquiry further.

The works of Scheffer et. al.  \cite{Scheffer1,Scheffer2} also demonstrate that mutation dynamics may lead to coexistence of many species within a single niche. We have demonstrated here that this principle holds generically for every trait that admits a quadratic fitness maximum. The details of \cite{Scheffer1,Scheffer2}  model, like directed mutations, nonlinear (density dependent) fitness function, interaction kernels and the effect of natural enemies  are irrelevant to this phenomena.

The neutral selection concept was presented here in the ecological
context, but all the above ideas may also be relevant, mutatis
mutandis, to other scenarios where the neutral dynamics seems to
emerge from system that admit strong competition, from  Kimura's
neutral theory of molecular evolution~\cite{kimura} to
macroevolutionary dynamics distributions~\cite{plos_one_maruvka}.
Different values of $\nu$ may be interpreted, accordingly,  as
different species (as we did here), different phenotypes,
quasi-species (as in Eigen-Schuster  model~\cite{eigen}) and so on.

Finally, our results hint at the possibility of a more radical viewpoint.  Our analysis  suggests that the
relevant grouping of individuals within a community is according to
their traits (here we consider only one trait, $\nu$, but the
extinction of the theory to a multi-dimensional fitness landscape is
trivial). One may wonder about the ratio between   inter-species and
intra-specific trait variations~\cite{ClarkScience}. If a typical
species is made of groups with a wide range of, say, $\nu$'s, and
other species admit similar values of this trait, the concept of
species itself may become a ``red herring" with respect to the
analysis of community dynamics.

\section*{Methods}

\subsection*{Neutral dynamics with environmental stochasticity}

 Let us consider the standard Hubbell-Kimura zero-sum dynamics, where in turn pairs of individuals are chosen, with one being replaced by the offspring of the other.  However, instead of the selected pair having equal probability of dying, the individual belonging to the less fit species has a higher probability of expiring. The species fitnesses are chosen at random and updated with some average rate. This model can be analyzed, leading to the Fokker-Planck equation for the distribution of species sizes, $P(m)$:
\begin{equation}
\frac{\partial P(m,t)}{\partial t} = S \frac{\partial^2}{\partial m^2} \left( m^2 P(m,t)\right) + \frac{\sigma^2}{2} \frac{\partial^2}{\partial m^2} \left(m P(m,t)\right) + \mu \frac{\partial }{\partial m}\left(m P(m,t)\right)
\end{equation}
The parameter $S$ measures the effect of environmental noise, which is proportional to the variance of the fitness distribution from which the species fitnesses are drawn and also to the typical persistence time, i.e., the time scale needed for a substantial change of the environmental conditions, i.e. the species' fitnesses.  When $S \to 0$, the standard neutral model is recovered.  The equilibrium solution of the Fokker-Planck equation is given in Eq. (\ref{ssnoise}) above.

\subsection*{Fixed Landscape Simulations}

 We have simulated the basic haploid Wright-Fisher evolution  model with a linear or  quadratic fitness maximum.  Each successive generation in this model is constructed by drawing parents from the current population with probability
\begin{equation}
P_i = \frac{f(x_i)}{\sum_j f(x_j)}
\end{equation}
where each individual $i$ is characterized by the trait $x_i$.  The fitness function $f(x)$ is given by
\begin{equation}
f(x) = e^{-s x^2}
\end{equation}
where $s$ is the selection strength.  For convenience, we work with a discrete fitness space labelled by the coordinate $x$, with unbiased mutational steps of $\Delta x \pm 1$.   We can think of the quantitative trait $x$ as arising from a large number of genes, each of which has two alleles, one of which contributes  to the trait and the other doesn't.  The individual has optimal fitness when half the alleles are contributing, which we call $x=0$.  Each haplotype is considered a separate species, so that there are no recurrent mutations.  This model is equivalent to a model with a continuous fitness space, where every value of $x$ constitutes a separate species, since if the contribution of the different genes has some small variation, every haplotype will have a slightly different fitness.  We have also simulated such a model, with equivalent results. 

 All SAD plots are constructed using a modified logarithmic binning, wherein the small $m$ data ($m\le 10$) are plotting without binning, and the data for larger $m$ is binned logarithmically. The Preston plots are computed such that octave bin $i$ counts the number of species of size $2^{i-1} < m \le 2^i$. Thus the data at the bin boundaries is {\em not} shared between bins, as in the original Preston prescription~\cite{Preston}.  Thus, the $0$th octave always shows a peak, as the number of spacies of size $1$ always exceeds that of size 2, a fact which is obscured in the sharing prescription~\cite{Williamson}

The results of the quadratic fitness landscape were contrasted to the Wright-Fisher model with a linear landscape.  Here the state $x=0$ has maximum fitness, and the fitness function is
\begin{equation}
f(x)=e^{-sx}
\end{equation}
Again we consider the limit of a large number of genes, so that there are no recurrent mutations, and all mutations from $x=0$  are disadvantageous.

To add environmental noise to the quadratic landscape model, we denote by $x_0$ the maximum of the fitness landscape, so that the fitness function now reads
\begin{equation}
f(x) = e^{-s(x-x_0)^2}
\end{equation}
We let $x_0$ vary randomly in time, updating it each generation by
\begin{equation}
x_0(t+1) = (1 - 1/\tau) x_0(t) + \delta \eta
\end{equation}
where $\eta$ is a Gaussian random variable with mean 0 and variance 1, and $\tau=200$ is the correlation time of the noise.  Here $\delta\sqrt{\tau}$ parameterizes the noise strength $S$.

We contrast this to a variant Hubbell model with environmental noise. Here, each {\em species} is assigned a random fitness, which is updated after
$\tau N$ competitions, where we have chosen $\tau=1$. In each competition, a pair of individuals, $(i,j)$, is chosen, and one of the two is replaced by the offspring of the other. The chance of $i$ winning is $f_i/(f_i + f_j)$.  The random fitnesses are drawn from a uniform distribution between $1\pm \delta$.  In addition, there is a probability $\mu$ for the offspring to mutate, giving rise to a new species.

\subsection*{Hamilton-May simulation}
Our second basic model is a  variant~\cite{hm,hm1} of the HM
process with explicit evolutionary dynamics (mutations).
We consider $L$ islands with a maximum carrying capacity of $N$
individuals on each island after the competition step. Each generation
starts with $m_i \leq N$ individuals on the $i$-th island. Every
individual produces a random number of offspring, where the number $q$ of offspring is an integer drawn for each given individual from a
Poisson distribution with average ${\bar{q}}$. All our
simulations were carried out with $N = 20, \ L = 500, \ {\bar{q}} =
2$.

Next comes the dispersal step: each individual stays at home with
probability $1-\nu$ and emigrates with probability $\nu$. An
emigrant survives the trip with probability $p$ (so $1-p$ is the
``migration cost'') and may end its voyage at any other island with
equal probability. In our simulations, we used $p=0.3$.

After the migration step the individuals on each overpopulated island compete among their island-mates, due to the limited carrying capacity of each island.  This is implemented by choosing randomly $N$ inhabitants to survive on
each island where there are currently more than $N$ individuals.  Underpopulated  islands are unaffected by the competition.

Finally, a local catastrophe occurs  on every given island with probability $X = 10\%$  and
kills all the local inhabitants. The cycle then starts again.

The chance of an individual to migrate is an inherited property, but
every now and then (1 out of 100 births) an offspring mutates from the $\nu$ of its father
to $\nu' = \nu + \delta \nu$.  The method of randomly choosing $\nu'$  is
subject to two restrictions: first we would like $0<\nu'<1$, and
second we would like to have no bias, i.e., on average the mutation
does not raise or lower $\nu$. To achieve that we have chosen $\nu'$
using the beta distribution, $\textrm{Beta}(\alpha,\beta(\nu))$. As the mean of the beta distribution is $\alpha/(\alpha+\beta)$, to preserve the mean
we choose $\beta(\nu) = \alpha(1/\nu -1)$.  The parameter $\alpha$ characterizes the width of the distribution, with the variance given (for our chosen $\beta$) by
\begin{equation}
\textrm{Var}(\nu') =  \frac{(1-\nu)\nu^2}{\alpha+\nu}
\end{equation}
which decreases monotonically with $\alpha$, and of course vanishes as $\nu$ approaches $0$ or $1$.  As with the quadratic fitness model, the fitting exponent $\beta$ of the generalized Fisher log-series, Eq. (\ref{gfls}), (not to be confused with the $\beta$ parameter of the beta distribution), approaches unity for large values of the $\alpha$ parameter.

\section*{Acknowledgments}
This work  was supported by the Israeli Ministry of science
TASHTIOT program and by the Israeli Science Foundation BIKURA grant
no. 1026/11.

\bibliography{bibfile}

\renewcommand{\thefigure}{S\arabic{figure}}
\setcounter{figure}{0}

\section*{Figure Legends}

\section*{Supporting Information}

\begin{figure}
\begin{center}
\includegraphics[width=0.6\textwidth]{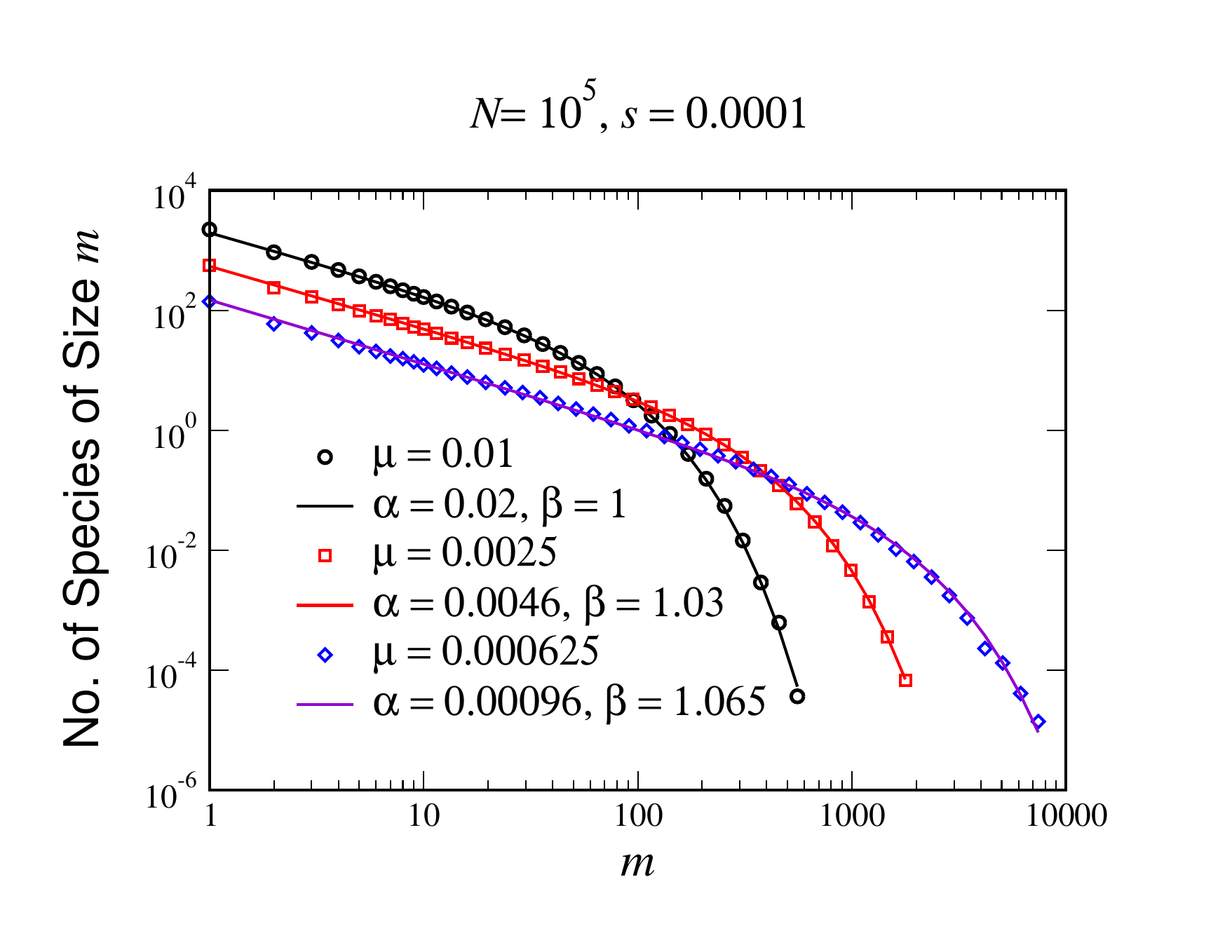}
\end{center}
\caption{The time-averaged SAD for the quadratic landscape for different mutation rates $\mu$'s. The selection parameter is $s=0.0001$. The other parameters are as in Fig. \ref{figmanysSAD}. Note that $\mu$ most strongly impacts on the $\alpha$ parameter of the generalized Fisher log-series, Eq. (\ref{gfls}).}
\label{figmanymu}
\end{figure}

\subsection*{Fig. \ref{figmanymu}: $\mu$ Dependence of SAD for Quadratic Landscape}

\begin{figure}
\begin{center}
\includegraphics[width=0.6\textwidth]{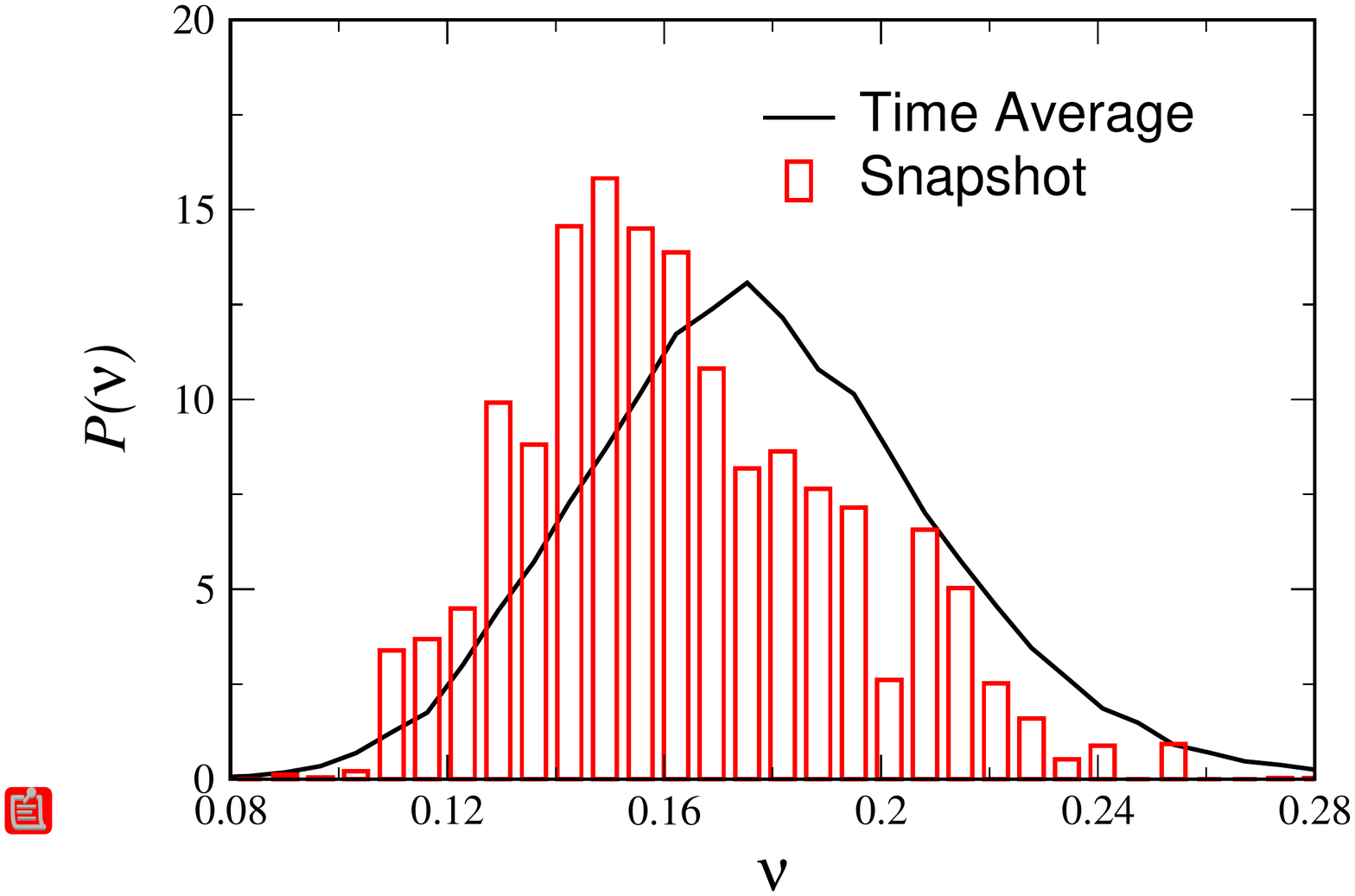}
\end{center}
\caption{The time-averaged distribution of $\nu$ in the Hamilton-May showing the effects of selection. Also shown is a snapshot of the distribution, showing the large fluctuations in the distribution due to environmental noise.  Note that in the noise-free quadratic landscape model, the snapshot is essentially indistinguishable from the time-average for the same metapopulation size.}
\label{fighmnuhist}
\end{figure}

\subsection*{Fig. \ref{fighmnuhist}: Distribution of $\nu$ in Hamilton-May Model}

\begin{figure}
\begin{center}
\includegraphics[width=0.6\textwidth]{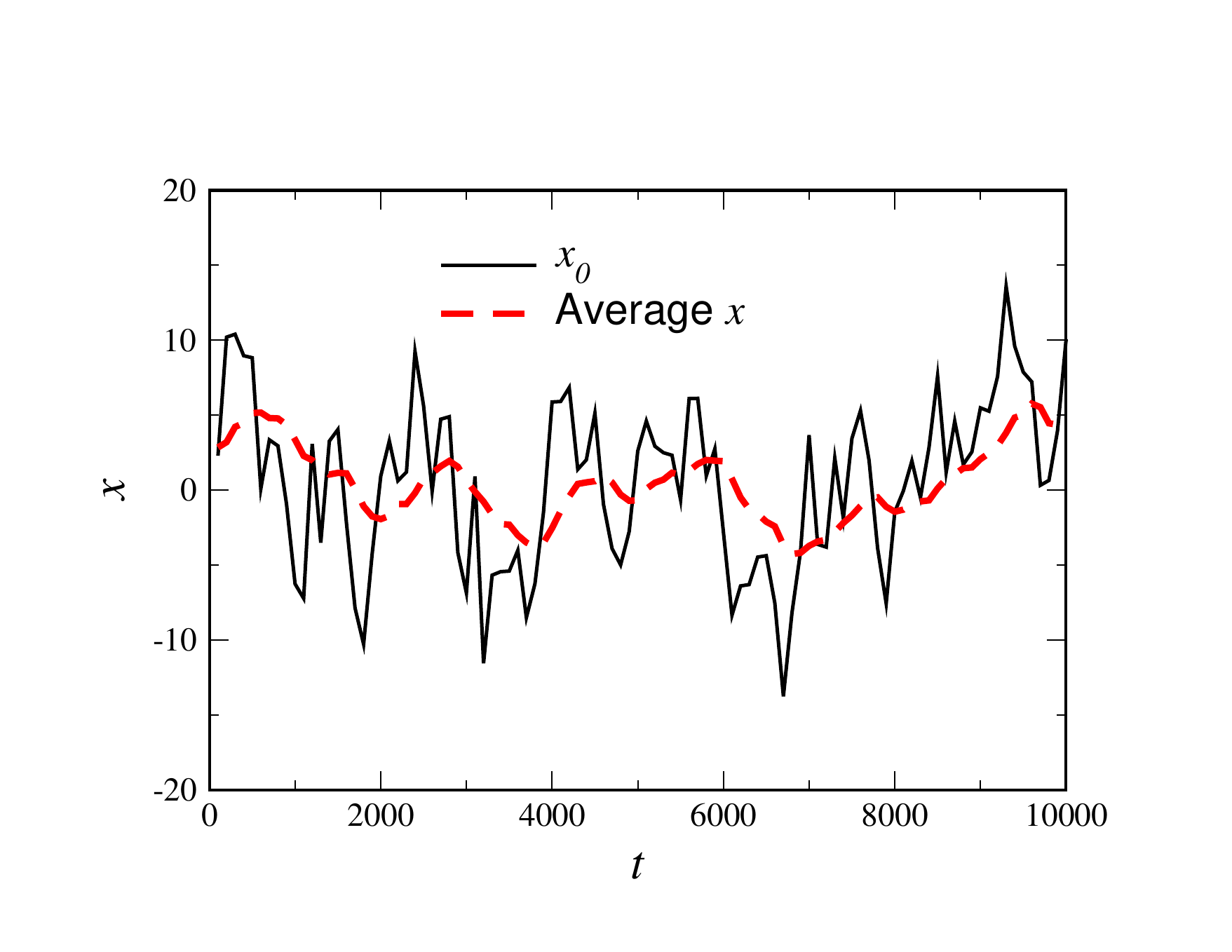}
\end{center}
\caption{The position of the fitness maximum as a function of time for the quadratic fitness model with environmental noise whose SAD is presented in Fig. \ref{figquadnoiseSAD}.  Also shown in the mean $x$ of the population as a function of time.}
\label{figxvst}
\end{figure}

\subsection*{Fig. \ref{figxvst}: Population Average $x(t)$ and Fitness Maximum $x_0(t)$ for the Quadratic Landscape with Environmental Noise}

\begin{figure}
\begin{center}
\includegraphics[width=0.6\textwidth]{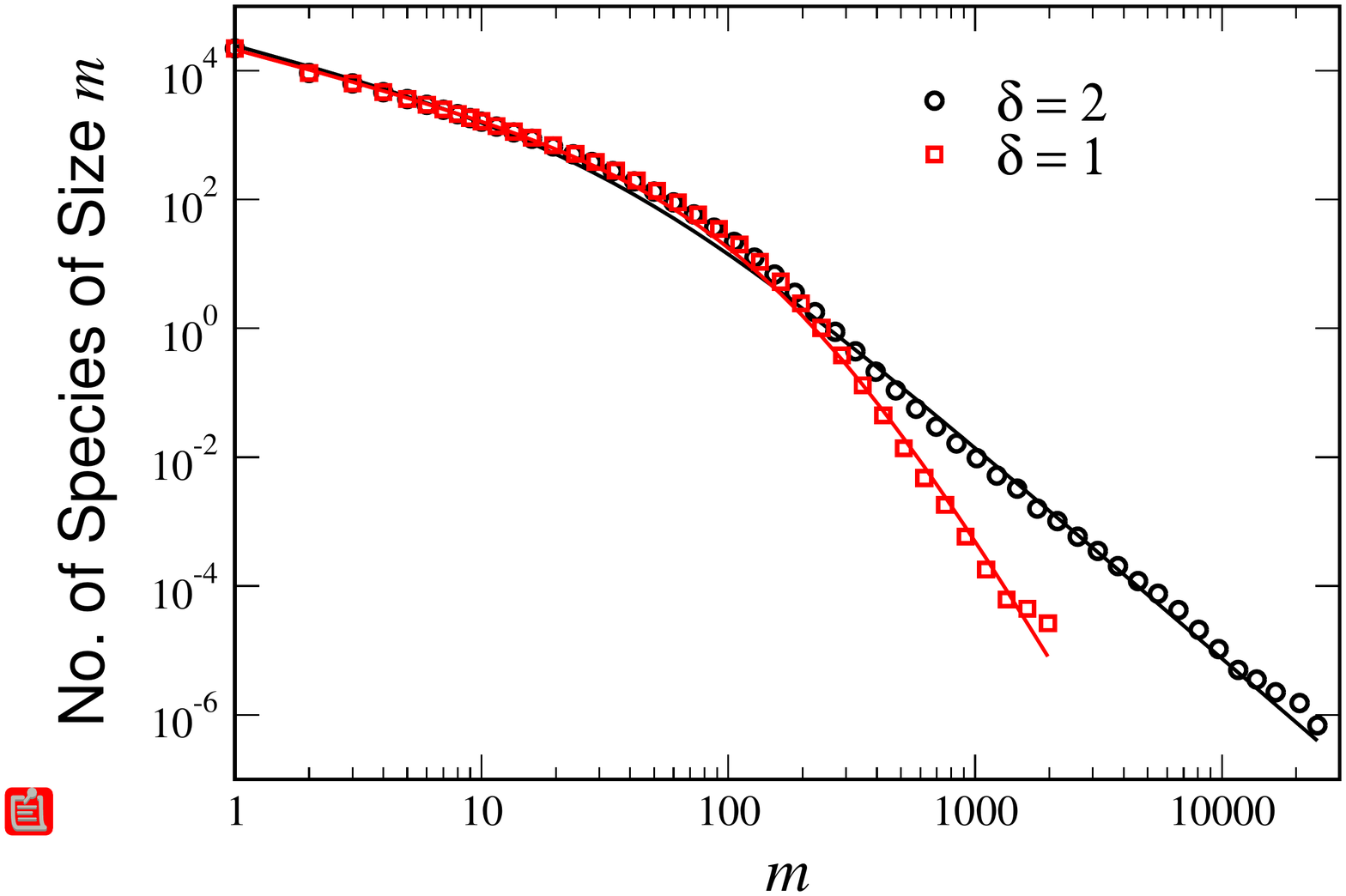}
\end{center}
\caption{Time-averaged SAD for the quadratic landscape model with large environmental noise ($\delta=1$ and 2, corresponding to twice and four times the width as for the moderate noise case presented in Fig. \ref{figquadnoiseSAD}.  The other parameters are unchanged.}
\label{figquadnoises}
\end{figure}

\subsection*{Fig. \ref{figquadnoises}: SAD for Quadratic Landscape with Large Environmental Noise}

\begin{figure}
\begin{center}
\includegraphics[width=0.6\textwidth]{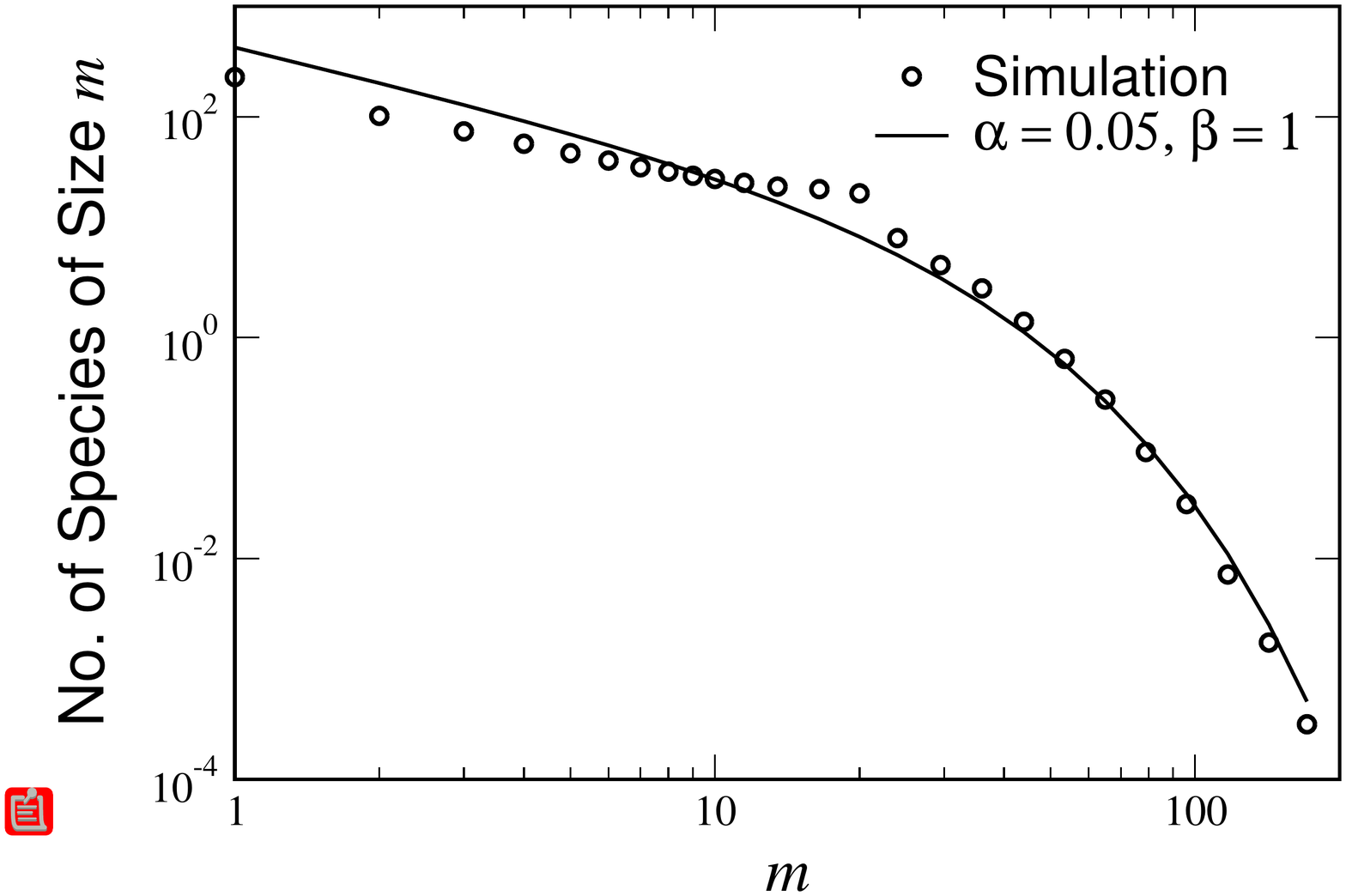}
\end{center}
\caption{Time-averaged SAD for the Hamilton-May  model without local catastrophes.  All other parameters are the same as for Fig. \ref{HMSAD}. Notice the peak at $m=20$, corresponding to the capturing of an island by a single species, and the subsequent inappropriateness  of the Fisher log-series fit.}
\label{fighm0x}
\end{figure}

\subsection*{Fig. \ref{fighm0x}: SAD for Hamilton-May without Local Catastrophes}

\end{document}